# THE ENHANCED DERIVED VECTOR SPACE APPROACH TO DOMAIN DECOMPOSITION METHODS

By
Ismael Herrera-Revilla,
Instituto de Geofísica
Universidad Nacional Autónoma de México (UNAM)
Email: iherrera@unam.mx

**Abstract**

*Standard approaches to domain decomposition methods (DDM) are uncapable of producing block-diagonal system matrices. The derived-vector-space (DVS), approach to DDM, introduced in 2013, overcomes this limitation. However, the DVS approach in its original form was applicable to a relatively narrow class of problems because it required building a special matrix, whose construction is frequently impossible. In this paper, an enhanced formulation of DVS is presented, which does not require the construction of a special matrix and is applicable to any linear problem.*

Equation Section 1

## 1.- INTRODUCTION

Computation is the innovation introduced in the XX Century that has most extensively changed human life.

Besides, computation is also a fundamental tool for advancing and using pure and applied science.

Today it is recognized that science and engineering stand in three columns [1]:

i) Theoretical Science,

ii) Experimental Science, and

iii) Computational Science.

Scientific-Behavior-Prediction is a fundamental component of Computational Science. Its main purpose is predicting the behavior of Nature and systems of human interest. In the case of Physical Sciences, Newton in the XVII Century showed that knowledge alone does not have the capacity of predicting behavior. For that purpose, it is required to integrate knowledge into mathematical models.

The mathematical modeling approach that Newton introduced in the XVII Century was a tremendous step forward, but its fruits remained limited for three centuries. This waiting period that went from the XVII to the XX Centuries was necessary for developing fully and deeply Newton's legacy. But, furthermore, by now it is clear that it was also necessary to wait for the computer's advent, because scientific behavior's prediction requires a tremendous number of arithmetic calculations that only computers and supercomputers have made feasible.

The fundamental mathematical models of pure and applied science are constituted by systems of partial differential equations. For physical systems, when the system is microscopical, Schrödinger Equation; and, for macroscopical systems, the counterpart of Schrödinger Equation is supplied by Herrera's axiomatic approach (see, [2]). All this, points to the fact that the treatment of partial differential equations -and more specifically, the solution of well-posed problems for such equations- is the key for applying computers to Scientific Prediction.

The solutions of such a kind of equations are functions that are defined at each point of the physical space. In order to apply computers to such problems, it is necessary to reduce that infinite number to a finite one and this is accomplished by means of discretization processes.

The first viable computer was constructed at the end of Second World War, in 1945, at the middle of the XX Century, and the pioneers of mathematical and computational prediction with great foresight devoted the second-half of that century, to developing the mathematical tools required for harnessing the future computers' capacity to serve Scientific Prediction. Therefore, the applied mathematical activity of the second-half of that century, world-wide, was focused on numerical methods for partial differential equations.

Parallel computing is outstanding among the new computational tools, especially at present when further increases in hardware speed apparently have reached insurmountable barriers. This kind of computation offers an attractive alternative for increasing processing-velocity and the international community has been working on it for a significant time-period.

Reckoning that *domain decomposition methods (DDM, or simply DD)* is the most efficient manner of solving in parallel partial differential equations was an important step forward and in 1988, the *DDM* or *DD organization* was founded. Since then, through its periodical and general activities this organization has established a body of knowledge, procedures, and frameworks that at present are generally accepted *canons* for most of the activity on domain decomposition methods [3-21].

All this has been very valuable. However, shortcomings are present as well. One, especially significant, is that *canonical approaches* produce system-matrices that are not block-diagonal. Due to this feature, they have to solve a coarse problem before finishing the numerical processing of problems they deal with (the summary of [21] explains this inconvenience with some detail).

In 2002, I hosted the *"Fourteenth International Conference on Domain Decomposition Methods"* [22]. Sometime after, we (I, together with my research group) began to develop a *non-canonical* approach to domain decomposition methods [23-31]. This methodology is known as the *derived-vector-space (DVS) approach* to *domain-decomposition-methods (DVS-DDM)*. A significant property of *DVS-DDM* is that contrary to *canonical approaches*, it yields block-diagonal system-matrices whenever it is applied.

At present, there are two versions of the DVS approach to DDM, here referred to as *DVS* and *'Enhanced-DVS'*, respectively. All the published material thus far refers to *DVS*, and the purpose of this article is presenting a systematic exposition of the *Enhanced-DVS (EDVS)*.

The goal of *EDVS*, and that of *DVS* (see, [207, 211]) as well, is supplying a tool for transforming numerical formulations suitable for treating partial differential equations sequentially, into formulations adequate for treating them in parallel.

Usually, when a problem is treated sequentially a system of nodes -the *"original-nodes"*- is introduced. Thus, the idea of the present article, is taking such nodes as the starting-point for constructing an *algebraic-venue* that is adequate for treating the same problem, but in parallel instead. Thus, the construction proceeds as follows, Sections 2 and 3, discuss *original-nodes*, while Section 4 discusses *derived-nodes*. Functions defined on finite sets yield finite-dimensional vector-spaces. In this manner is produced the *derived-vector space* (*DVS*), discussed in Section 5. Such a space is decomposed into *continuous* and *zero-average* subspaces in Section 6, while in Section 7 the *immersion* of the *original-vectors* space into the *derived-vector space* is explained. Section 8 is devoted to the immersion of matrices and, there the concept of *dual-matrices* is introduced. The explicit expressions for *dual-matrices* are given in Section 9. The "EDVS Parallel Formulation" of well-posed boundary value problems is given in Section 10, while "General Schur-Complement Decompositions" is given in 11. Sections 12 and 13 are Application of the EDVS-DDM Formulation and Conclusions, respectively.

Equation Section 2

## 2.- THE ORIGINAL NODES

We consider an open and bounded domain $\Omega \subset \mathbb{R}^n$ and a finite set of points $\hat{X} \subset \Omega$. When $p \in \hat{X} \subset \Omega$, then $p$ is said to be an *original-node*. It is assumed that the cardinality of $\hat{X}$ is $N > 0$.

Next, we consider a *'domain decomposition'* of $\Omega$; by this, we mean a family $\{\Omega_1, \ldots, \Omega_E\}$ of subdomains of $\Omega$ such that, for each $\alpha = 1, \ldots, E$, $\Omega_\alpha$ is a bounded domain and the following properties are satisfied:

$i).\ \Omega_\alpha \subset \Omega;\ \alpha = 1,...,\mathrm{E}$

$ii).\ \Omega_\alpha \cap \Omega_\beta = \varnothing, \forall \alpha \neq \beta$

$iii).\ \partial_{\mathrm{I}}\Omega \equiv \Omega - \bigcup_{\alpha=1}^{\mathrm{E}} \Omega_\alpha = \bigcup_{\alpha\neq\beta} \left( \bar{\Omega}_\alpha \cap \bar{\Omega}_\beta \right)$, and

$iv).\ \partial\Omega = \bigcup_{\alpha=1}^{\mathrm{E}} \partial\Omega_\alpha - \partial_{\mathrm{I}}\Omega$

Equation Section 3

## 3.- CLASSIFICATION OF ORIGINAL NODES

We define

$$\widehat{\mathrm{X}}^\alpha \equiv \left\{ p \in \widehat{\mathrm{X}} \middle| p \in \bar{\Omega}_\alpha \right\},\ \alpha = 1,...,E \tag{3.1}$$

Clearly,

$$\widehat{\mathrm{X}} = \widehat{\mathrm{X}}^1 \cup ... \cup \widehat{\mathrm{X}}^E = \bigcup_{\alpha=1}^{\alpha=E} \widehat{\mathrm{X}}^\alpha \tag{3.2}$$

However, generally

$$\widehat{\mathrm{X}}^\alpha \cap \widehat{\mathrm{X}}^\beta \neq \varnothing \ when\ \alpha \neq \beta \tag{3.3}$$

Let $p \in \widehat{\mathrm{X}}$ be an *original node*, then define

$$\delta_\alpha(p) \equiv \begin{cases} 1, \text{if } p \in \widehat{\mathrm{X}}^\alpha \\ 0, \text{if } p \notin \widehat{\mathrm{X}}^\alpha \end{cases} \tag{3.4}$$

The *multiplicity of* $p$, $m(p)$, is defined to be

$$m(p) \equiv \sum_{\alpha=1}^{\alpha=\mathrm{E}} \delta_\alpha(p) \tag{3.5}$$

The subsets $\widehat{\mathrm{I}} \subset \widehat{\mathrm{X}}$ and $\widehat{\Gamma} \subset \widehat{\mathrm{X}}$ are defined by:

$$\widehat{\mathrm{I}} \equiv \left\{ p \in \widehat{\mathrm{X}} \middle| m(p) = 1 \right\} \text{and}\ \widehat{\Gamma} \equiv \left\{ p \in \widehat{\mathrm{X}} \middle| m(p) > 1 \right\} \tag{3.6}$$

Any $p \in \hat{\mathrm{X}}$ is said to *'interior'* when $p \in \hat{\mathrm{I}}$ and *'interphase'* when $p \in \hat{\Gamma}$. In words, $p \in \hat{\mathrm{X}}$ is said to be *interior* when its *multiplicity* is one, and *'interphase'* when its *multiplicity* is greater than one.

It can be verified that the pair $\{\hat{\mathrm{I}}, \hat{\Gamma}\}$ *decomposes* $\hat{X}$; i.e., that (see the Appendix on set theory, for a general definition of a set decomposition):

$$\hat{X} = \hat{\mathrm{I}} \cup \hat{\Gamma} \text{ and } \varnothing = \hat{\mathrm{I}} \cap \hat{\Gamma} \qquad (3.7)$$

Equation Section 4

## 4.- DERIVED-NODES AND ITS CLASSIFICATION

The so-called, *divide and conquer* strategy of *domain decomposition methods (DDM)*, after introducing a domain decomposition of $\Omega$, requires subsequently establishing a one-to-one mapping of the subdomains onto the available processor cores. Ideally, with the purpose of processing in different processors, the nodes contained in different subdomains. However, this is impossible when $\hat{\Gamma} \neq \varnothing$.

In *canonical methods* the condition $\hat{\Gamma} = \varnothing$ is never satisfied. As a consequence of this fact, the system matrix is not block-diagonal in such methods. The construction that follows is relevant for overcoming that shortcoming.

To start with, we define the following sets of pairs:

$$\mathrm{X}^{\alpha} \equiv \left\{ (p, \alpha) \middle| \, p \in \hat{\mathrm{X}}^{\alpha} \right\}, \; \alpha = 1, \ldots, E \qquad (4.1)$$

Recall that $E$ is the number of subdomains contained by the *domain decomposition*. Also notice that according Eq.(4.1), the members of $\mathrm{X}^{\alpha}$ are by pairs, the first one is any $p \in \hat{\mathrm{X}}$ and the second is $\alpha \in \{1, \ldots, E\}$ such that $p \in \hat{\mathrm{X}}^{\alpha}$.

Furthermore, the above definition of $\mathrm{X}^{\alpha}$, for $\alpha \in \{1, \ldots, E\}$, implies

$$\mathrm{X}^{\alpha} \cap \mathrm{X}^{\beta} = \varnothing, \; \forall \alpha \neq \beta \qquad (4.2)$$

The *whole set of derived-nodes* is defined to be

$$\mathrm{X} \equiv \mathrm{X}^1 \cup \ldots \cup \mathrm{X}^E = \bigcup_{\alpha=1}^{\alpha=E} \mathrm{X}^\alpha \qquad (4.3)$$

Eqs. (4.2) and (4.3) together imply that the family of subsets $\left\{ \mathrm{X}^1, \ldots, \mathrm{X}^E \right\}$ decomposes $\mathrm{X}$. We also observe that an equivalent definition, alternative to Eq. (4.3), is:

$$\mathrm{X} \equiv \left\{ (p, \alpha) \middle| \, p \in \mathrm{X}^\alpha \right\} \qquad (4.4)$$

Corresponding to each *original node*, $p \in \hat{\mathrm{X}}$, we define the set

$$\mathrm{Z}(p) \equiv \left\{ (p, \alpha) \middle| (p, \alpha) \in \mathrm{X} \right\} \qquad (4.5)$$

of derived-nodes. We refer to $\mathrm{Z}(p)$ as the *descendants of* $p$. Thereby, we observe that the *cardinality* of $\mathrm{Z}(p)$ is $m(p)$, as defined by Eq.(3.5). We also notice that

$$\mathrm{X} = \bigcup_{p \in \hat{\mathrm{X}}} \mathrm{Z}(p) \text{ and } \mathrm{Z}(p) \cap \mathrm{Z}(q) = \varnothing, \text{ whenever } p \neq q \qquad (4.6)$$

Hence, the family of subsets $\left\{ \mathrm{Z}(1), \ldots, \mathrm{Z}(\mathrm{N}) \right\}$ decomposes $\mathrm{X}$.

The following classification applies to *derived-nodes*:

$$\begin{aligned}
&\mathrm{X}^\alpha \equiv \left\{ (p, \alpha) \middle| \, p \in \bar{\Omega}_\alpha \right\}, \ \alpha = 1, \ldots, E \\
&\mathrm{I} \ \textit{internal-nodes;} \ (p, \alpha) \in \mathrm{I} \Leftrightarrow m(p) = 1 \\
&\Gamma \ \textit{interface-nodes;} \ (p, \alpha) \in \mathrm{I} \Leftrightarrow m(p) > 1 \\
&\pi \subset \Gamma \ \textit{primal-nodes} \\
&\Delta \subset \Gamma \ \textit{dual-nodes} \\
&\Pi \equiv \mathrm{I} \cup \pi \\
&\Sigma \equiv \mathrm{I} \cup \Delta
\end{aligned} \qquad (4.7)$$

Then:

$$\left\{ \mathrm{I}, \Delta, \pi \right\} \ \textit{decomposes} \ \mathrm{X} \qquad (4.8)$$

Also,

$$\{\mathrm{I},\Gamma\}\ decompose\ \mathrm{X}$$
$$\{\Pi,\Delta\}\ decompose\ \mathrm{X} \quad (4.9)$$
$$\{\Sigma,\pi\}\ decompose\ \mathrm{X}$$

While

$$\{\Delta,\pi\}\ decompose\ \Gamma$$
$$\{\mathrm{I},\pi\}\ decompose\ \Pi \quad (4.10)$$
$$\{\mathrm{I},\Delta\}\ decompose\ \Sigma$$

Equation Section 5

## 5.- THE VECTOR SPACES $\widehat{W}$ AND $W$

In this Section, we introduce two function spaces and several inner-products that will be used in the sequel.

### 5.1 The original-vectors space $\widehat{W}$

To start with, we consider functions of the form $\underline{\hat{u}}:\widehat{\mathrm{X}}\to\mathrm{Y}$, to be called *original-vectors*; they are defined on the *original-nodes* and take values on the counter-domain $\mathrm{Y}$, which is assumed to be a linear space. The set of all *original-vectors* constitutes a linear space, denoted by $\widehat{W}$, with respect to the usual definitions of sum and multiplication by scalars (which here are real numbers), for functions that take values on a linear space. Moreover, when $\underline{\hat{u}}\in\widehat{W}$ and $p\in\widehat{\mathrm{X}}$, then $\underline{\hat{u}}(p)\in\mathrm{Y}$ is the value of $\underline{\hat{u}}$ at $p$.

### 5.2 The derived-vectors space $W$.

We consider also functions of the form $\underline{u}:\mathrm{X}\to\mathrm{Y}$, to be called *derived-vectors*; they are defined on *derived-nodes* and take values on the linear space $\mathrm{Y}$. With similar conventions to those adopted in the case of $W$, the set of all *derived-vectors* also constitutes a linear space, denoted by $W$. Moreover, when $\underline{u}\in W$ and $(p,\alpha)\in\mathrm{X}$, then $\underline{u}(p,\alpha)\in\mathrm{Y}$ is the value of $\underline{u}$ at $(p,\alpha)$.

### 5.3 The inner-products

In what follows several real-valued inner-products will be considered. Two such inner-products are denoted by $\langle \bullet,\bullet \rangle_Y$ and $\langle\langle \bullet,\bullet \rangle\rangle_Y$, respectively. These are defined in $\mathrm{Y}$, which is assumed to be a finite dimensional Hilbert Space. Besides, they are related by the condition

$$\langle \bullet,\bullet \rangle_{\mathrm{Y}} = m(p)\langle\langle \bullet,\bullet \rangle\rangle_{\mathrm{Y}} \tag{5.1}$$

When $a \in \mathrm{Y}$ and $b \in \mathrm{Y}$, we write $\langle a,b \rangle_Y$ and $\langle\langle a,b \rangle\rangle_Y$ for their values.

Other inner-products that will be used are $\langle \bullet,\bullet \rangle$, which is defined in $\widehat{W}$, and $\langle\langle \bullet,\bullet \rangle\rangle$ defined in $W$. Moreover:

$$\langle \underline{\hat{u}},\underline{\hat{v}} \rangle \equiv \sum_{p\in\hat{\mathrm{X}}} \langle \underline{\hat{u}}(p),\underline{\hat{v}}(p) \rangle_{\mathrm{Y}},\ \forall \underline{\hat{u}},\underline{\hat{v}} \in \widehat{W} \tag{5.2}$$

and

$$\langle\langle \underline{u},\underline{v} \rangle\rangle \equiv \sum_{(p,\alpha)\in\mathrm{X}} \langle\langle \underline{u}(p,\alpha),\underline{v}(p,\alpha) \rangle\rangle_{\mathrm{Y}},\ \forall \underline{u},\underline{v} \in W \tag{5.3}$$

This definition can also be written as

$$\langle\langle \underline{u},\underline{v} \rangle\rangle \equiv \sum_{p\in\hat{\mathrm{X}}} \sum_{(p,\alpha)\in Z(p)} \langle\langle \underline{u}(p,\alpha),\underline{v}(p,\alpha) \rangle\rangle_{\mathrm{Y}},\ \forall \underline{u},\underline{v} \in W \tag{5.4}$$

The equivalence of these latter two equations is due to the fact that the family of subsets $\{\mathrm{Z}(1),...,\mathrm{Z}(\mathrm{N})\}$ decomposes $\mathrm{X}$, as was seen in Section 4.

When the linear spaces $\widehat{W}$ and $W$ are complemented with the inner-products just introduced, each one them becomes a finite dimensional Hilbert Space.

Equation Section 6

## 6.- CONTINUOUS AND ZERO-AVERAGE SUBSPACES

**DEFINITION 6.1.**

A *derived-vector,* $\underline{u} \in W$, is said to be *'continuous at* $p \in \hat{X}$*'* when $\underline{u}(p,\alpha)$ takes a unique value for very $(p,\alpha) \in Z(p)$. When $\underline{u} \in W$ is continuous at every $p \in \hat{X}$, then it is said to be *'continuous'*.

**REMARKS**

1. The subset of *continuous derived-vectors* constitutes a linear subspace of $W$ that will be denoted by $W_I$;

2. When $\underline{u} \in W$ is *continuous at* $p \in \hat{X}$, then

$$\sum_{(p,\alpha)\in Z(p)} \underline{u}(p,\alpha) = m(p)\underline{u}(p,\beta), \forall (p,\beta) \in Z(p) \tag{6.1}$$

**DEFINITION 6.2.**

A *derived-vector,* $\underline{u} \in W$, is said to be *'zero-average at* $p \in \hat{X}$*'* when

$$\sum_{(p,\alpha)\in Z(p)} \underline{u}(p,\alpha) = 0 \tag{6.2}$$

When $\underline{u} \in W$, is *zero-average* at every $p \in \hat{X}$, then it is said to be *'zero-average'*. The subset of *zero-average derived-vectors* constitutes a linear subspace of $W$ that will be denoted by $W_{II}$.

**DEFINITION 6.3.**

For every $\underline{u} \in W$, the mappings $\underline{\underline{a}} : W \to W_I$ and $\underline{\underline{j}} : W \to W_{II}$ are defined respectively, by

$$\left(\underline{\underline{a}}\underline{u}\right)(p,\alpha) \equiv \frac{1}{m(p)} \sum_{(p,\beta)\in Z(p)} \underline{u}(p,\beta), \forall (p,\alpha) \in X \tag{6.3}$$

and

$$\underline{\underline{j}}\underline{u} \equiv \underline{u} - \underline{\underline{a}}\underline{u} \tag{6.4}$$

**REMARK.** Eq.(6.4), implies that every $\underline{u} \in W$, can be written as

$$\underline{u} = \underline{\underline{a}}\underline{u} + \underline{\underline{j}}\underline{u} \tag{6.5}$$

**THEOREM 6.1.**

The subspaces $W_I$ and $W_{II}$, are orthogonal subspaces of $W$, with respect to the inner product $\langle\langle \bullet, \bullet \rangle\rangle$.

<u>Proof</u>. Let $\underline{u} \in W_I$ and $\underline{v} \in W_{II}$. Then, by virtue of Eq.(5.4)

$$\langle\langle \underline{u}, \underline{v} \rangle\rangle = \sum_{p \in \widehat{X}} \sum_{(p,\alpha) \in Z(p)} \langle\langle \underline{u}(p,\alpha), \underline{v}(p,\alpha) \rangle\rangle_{Y}, \ \forall \underline{u}, \underline{v} \in W \tag{6.6}$$

Let be $(p,\beta) \in Z(p)$, where $p \in \widehat{X}$. Then

$$\sum_{(p,\alpha) \in Z(p)} \langle\langle \underline{u}(p,\alpha), \underline{v}(p,\alpha) \rangle\rangle_{Y} = \left\langle\left\langle \underline{u}(p,\beta), \sum_{(p,\alpha) \in Z(p)} \underline{v}(p,\alpha) \right\rangle\right\rangle_{Y} \tag{6.7}$$

Moreover,

$$\sum_{(p,\alpha) \in Z(p)} \underline{v}(p,\alpha) = 0 \tag{6.8}$$

because $\underline{v}$ is a *zero-average* function. Hence, $\langle\langle \underline{u}, \underline{v} \rangle\rangle = 0$, and the proof of Theorem 6.1 is complete.

This theorem, together with Eq.(6.5) implies that every $\underline{u} \in W$ be written uniquely as

$$\underline{u} = \underline{u}_I + \underline{u}_{II}, \text{ with } \underline{u}_I \in W_I \text{ and } \underline{u}_{II} \in W_{II} \tag{6.9}$$

with

$$\underline{u}_I \equiv \underline{\underline{a}}\underline{u} \text{ and } \underline{u}_{II} \equiv \underline{\underline{j}}\underline{u} \tag{6.10}$$

Equation Section 7

## 7.- THE IMERSION OF $\widehat{W}$ INTO $W_I$ AND DUALITY

In this Section, $\underline{\hat{u}}$ and $\underline{\hat{v}}$ stand for *original-vectors*, while $\underline{u}$ and $\underline{v}$ stand for *derived-vectors* that belong to $W_I$ .

**DEFINITION 7.1.**

The mapping $\underline{\underline{\hat{a}}}:\widehat{W} \to W_I$ is defined for every $\underline{\hat{u}} \in \widehat{W}$ by

$$\left(\underline{\underline{\hat{a}}}\underline{\hat{u}}\right)(p,\alpha) \equiv \underline{\hat{u}}(p), \forall (p,\alpha) \in X \tag{7.1}$$

Hence, when $\underline{u} = \underline{\underline{\hat{a}}}\underline{\hat{u}}$, then

$$\underline{u}(p,\alpha) = \underline{\hat{u}}(p), \forall (p,\alpha) \in X \tag{7.2}$$

**DEFINITION 7.2.** Define $\underline{\underline{\hat{a}}}^{-1}:W_I \to \widehat{W}$, for every $\underline{u} \in W_I$, by

$$\left(\underline{\underline{\hat{a}}}^{-1}\underline{u}\right)(p) = \frac{1}{m(p)} \sum_{(p,\alpha)\in Z(p)} \underline{u}(p,\alpha) =, \forall p \in \widehat{X} \tag{7.3}$$

**REMARK.** Definition 7.2 is effectively introduced as a definition. Therefore, thus far it does not have any relation with the inverse of $\underline{\underline{\hat{a}}}:\widehat{W} \to W_I$.

**LEMMA 7.1.**

For any $\underline{\hat{u}} \in \widehat{W}$, the condition $\underline{u} = \underline{\underline{\hat{a}}}\underline{\hat{u}}$ is satisfied, if and only if, for every $p \in \widehat{X}$,

$$\underline{\hat{u}}(p) = \underline{u}(p,\alpha), \forall (p,\alpha) \in Z(p) \tag{7.4}$$

Proof. Clearly $\underline{u} = \underline{\underline{\hat{a}}}\underline{\hat{u}}$ is equivalent to

$$\underline{\hat{u}}(p) = \underline{u}(p,\alpha), \forall (p,\alpha) \in X \tag{7.5}$$

by virtue of Definition 7.1. Therefore, $\underline{u} = \underline{\underline{\hat{a}}}\underline{\hat{u}}$ implies Eq.(7.4) for every $p \in \widehat{X}$, since $Z(p) \subset X$, for every $p \in \widehat{X}$. The converse follows from the fact that

$$X = \bigcup_{p\in\widehat{X}} Z(p) \tag{7.6}$$

as was shown in Eq. (4.6), Section 4.

**LEMMA 7.2.**

$$\sum_{(p,\alpha)\in Z(p)} \left(\underline{\underline{\widehat{a}}}\,\underline{\widehat{u}}\right)(p,\alpha) = m(p)\underline{\widehat{u}}(p) \tag{7.7}$$

Proof. It follows directly from Eq.(7.1).

**COROLLARY 7.1.**

$$\underline{u} = \underline{\underline{\widehat{a}}}\,\underline{\widehat{u}} \Leftrightarrow \left(\underline{\widehat{u}}(p) = \underline{u}(p,\alpha), \forall (p,\alpha) \in \mathrm{X}\right) \tag{7.8}$$

Proof. It follows from Eq.(7.7), because of the terms of the left-hand's addition of that equation are equal.

**COROLLARY 7.2.** The mapping $\underline{\underline{\widehat{a}}}^{-1} : W_I \to \widehat{W}$, as defined by Definition 7.2, is the inverse of $\underline{\underline{\widehat{a}}} : \widehat{W} \to W_I$.

Proof. It follows directly from Eq.(7.7).

**REMARK.** In view of the above results,

$$\underline{u} = \underline{\underline{\widehat{a}}}\,\underline{\widehat{u}} \Leftrightarrow \underline{\widehat{u}} = \underline{\underline{\widehat{a}}}^{-1}\underline{u} \tag{7.9}$$

**DEFINITION 7.3.** The vectors $\underline{u}$ and $\underline{\widehat{u}}$ are said to be *duals* of each other, when $\underline{u} = \underline{\underline{\widehat{a}}}\,\underline{\widehat{u}}$ or $\underline{\widehat{u}} = \underline{\underline{\widehat{a}}}^{-1}\underline{u}$. We use the notation $\underline{\widehat{u}} \sim \underline{u}$ to indicate that $\underline{u}$ and $\underline{\widehat{u}}$ are *duals* of each other.

**REMARK.** The relation $\bullet \sim \bullet$ is a symmetric relation.

**THEOREM 7.1.**

The condition $\underline{\widehat{u}} \sim \underline{u}$ is fulfilled, if and only if, the following condition is satisfied for every $p \in \widehat{\mathrm{X}}$,

$$\underline{\widehat{u}}(p) = \underline{u}(p,\alpha), \forall (p,\alpha) \in \mathrm{Z}(p) \tag{7.10}$$

Proof. The condition $\underline{\hat{u}} \sim \underline{u}$, is equivalent to the condition $\underline{u} = \underline{\underline{\hat{a}}}\underline{\hat{u}}$, and this in turn is equivalent to the assumption of this Theorem by virtue of Lemma 7.1.

**REMARK.** An alternative definition of the relation $\bullet \sim \bullet$, equivalent to Definition 7.3, is: '*we write* $\underline{\hat{u}} \sim \underline{u}$, *when Eq.(7.10) is fulfilled for every* $p \in \hat{\mathrm{X}}$'.

**THEOREM 7.2.** Using the notation and definitions adopted thus far, we have

$$\langle\langle \underline{u}, \underline{v} \rangle\rangle = \langle \underline{\hat{u}}, \underline{\hat{v}} \rangle, \text{ whenever } \underline{u} \sim \underline{\hat{u}} \text{ and } \underline{v} \sim \underline{\hat{v}} \tag{7.11}$$

Proof. Applying Eq.(5.4)

$$\langle\langle \underline{u}, \underline{v} \rangle\rangle = \sum_{p \in \hat{\mathrm{X}}} \sum_{(p,\alpha) \in Z(p)} \langle\langle \underline{u}(p,\alpha), \underline{v}(p,\alpha) \rangle\rangle_{\mathrm{Y}} \tag{7.12}$$

When $\underline{u} \sim \underline{\hat{u}}$ and $\underline{v} \sim \underline{\hat{v}}$, in view of Eq.(7.1), we have

$$\langle\langle \underline{u}, \underline{v} \rangle\rangle = \sum_{p \in \hat{\mathrm{X}}} m(p) \langle\langle \underline{\hat{u}}(p), \underline{\hat{v}}(p) \rangle\rangle_{\mathrm{Y}} \tag{7.13}$$

Applying in Eq.(5.1), it is seen that

$$\langle\langle \underline{u}, \underline{v} \rangle\rangle = \sum_{p \in \hat{\mathrm{X}}} \langle \underline{\hat{u}}(p), \underline{\hat{v}}(p) \rangle_{\mathrm{Y}} = \langle \underline{\hat{u}}, \underline{\hat{v}} \rangle \tag{7.14}$$

So, the proof of Theorem 7.2 is complete.

Equation Section 8

## 8.- DUALITY CONCEPTS FOR MATRICES

In this Section, $\underline{\underline{Q}}$ and $\underline{\underline{\hat{Q}}}$ stand for the linear transformations $\underline{\underline{\hat{Q}}}: \hat{W} \to \hat{W}$ and $\underline{\underline{Q}}: W_I \to W_I$, respectively.

**DEFINITION 8.1**

The linear transformations $\underline{\underline{\hat{Q}}}: \hat{W} \to \hat{W}$ and $\underline{\underline{Q}}: W_I \to W_I$, are said to be *duals of each other* when the pair of matrices $\underline{\underline{Q}}$ and $\underline{\underline{\hat{Q}}}$ are such that

$$\underline{\underline{Q}}\underline{u} \sim \underline{\underline{\hat{Q}}}\underline{\hat{u}}, \text{ whenever } \underline{u} \sim \underline{\hat{u}} \tag{8.1}$$

When $\underline{\underline{\hat{Q}}}: \hat{W} \to \hat{W}$ and $\underline{\underline{Q}}: W_I \to W_I$, are *duals of each other* the notations

$$\underline{\underline{Q}} \sim \widehat{\underline{\underline{Q}}} \text{ and } \widehat{\underline{\underline{Q}}} \sim \underline{\underline{Q}} \tag{8.2}$$

will be used.

We observe that the *duality* for matrices, establishes a symmetric relation between the matrices of the form $\widehat{\underline{\underline{Q}}} : \widehat{W} \to \widehat{W}$ and those of the form $\underline{\underline{Q}} : W_I \to W_I$. That is,

$$\underline{\underline{Q}} \sim \widehat{\underline{\underline{Q}}} \textit{ and } \widehat{\underline{\underline{Q}}} \sim \underline{\underline{Q}} \textit{ equivalent.} \tag{8.3}$$

Equation Section 9

## 9.- EXPLICIT EXPRESSION FOR THE DUAL MATRIX

Assume $\widehat{\underline{\underline{Q}}} \sim \underline{\underline{Q}}$ . Recall Eq. (8.1),

$$\underline{\underline{Q}}\underline{u} \sim \widehat{\underline{\underline{Q}}}\widehat{\underline{u}} \text{ whenever } \widehat{\underline{u}} \sim \underline{u} \tag{9.1}$$

Hence

$$\underline{\underline{Q}}\underline{u} = \widehat{\underline{\underline{a}}}\widehat{\underline{\underline{Q}}}\widehat{\underline{u}} = \widehat{\underline{\underline{a}}}\widehat{\underline{\underline{Q}}}\widehat{\underline{\underline{a}}}^{-1}\underline{u}, \forall \underline{u} \in W_I \tag{9.2}$$

Therefore,

$$\underline{\underline{Q}} \equiv \widehat{\underline{\underline{a}}}\widehat{\underline{\underline{Q}}}\widehat{\underline{\underline{a}}}^{-1} \tag{9.3}$$

Hence, the dual of any matrix $\underline{\underline{Q}} : W_I \to W_I$ is unique and, when $\widehat{\underline{\underline{Q}}} : \widehat{W} \to \widehat{W}$ is given, and the explicit expression of its *dual matrix* is supplied by Eq.(9.3). However, for applications it is useful to elaborate this equation a little further, as we do next.

Recall that $\widehat{\mathrm{I}} \subset \widehat{\mathrm{X}}$ and $\widehat{\Gamma} \subset \widehat{\mathrm{X}}$ decompose $\mathrm{X}$, while $\mathrm{I} \subset \mathrm{X}$ and $\Gamma \subset \mathrm{X}$ decompose $\mathrm{X}$. Hence:

$$\widehat{\mathrm{X}} = \widehat{\mathrm{I}} \cup \widehat{\Gamma} \text{ and } \varnothing = \widehat{\mathrm{I}} \cap \widehat{\Gamma} \tag{9.4}$$

and

$$\mathrm{X} = \mathrm{I} \cup \Gamma \text{ and } \varnothing = \mathrm{I} \cap \Gamma \tag{9.5}$$

Moreover, define

$$\widehat{W}_I^{\mathrm{I}} \equiv \widehat{W}_I\left(\widehat{\mathrm{I}}\right) \text{ and } \widehat{W}_I^{\Gamma} \equiv \widehat{W}_I\left(\widehat{\Gamma}\right) \tag{9.6}$$

together with

$$W_I^{\mathrm{I}} \equiv W_I(\mathrm{I}) \text{ and } W_I^{\Gamma} \equiv W_I(\Gamma) \quad (9.7)$$

Then,

$$\widehat{W}_I = \widehat{W}_I^{\mathrm{I}} \oplus \widehat{W}_I^{\Gamma} \quad (9.8)$$

and

$$W_I = W_I^{\mathrm{I}} \oplus W_I^{\Gamma} \quad (9.9)$$

All this leads to

$$\underline{\underline{\widehat{Q}}} = \begin{pmatrix} \underline{\underline{\widehat{Q}}}_{\mathrm{II}} & \underline{\underline{\widehat{Q}}}_{\mathrm{I}\Gamma} \\ \underline{\underline{\widehat{Q}}}_{\Gamma\mathrm{I}} & \underline{\underline{\widehat{Q}}}_{\Gamma\Gamma} \end{pmatrix} \text{ and } \underline{\underline{Q}} = \begin{pmatrix} \underline{\underline{Q}}_{\mathrm{II}} & \underline{\underline{Q}}_{\mathrm{I}\Gamma} \\ \underline{\underline{Q}}_{\Gamma\mathrm{I}} & \underline{\underline{Q}}_{\Gamma\Gamma} \end{pmatrix} \quad (9.10)$$

Here,

$$\underline{\underline{\widehat{Q}}}_{\mathrm{II}} : \widehat{W}_I^{\mathrm{I}} \to \widehat{W}_I^{\mathrm{I}},\ \underline{\underline{\widehat{Q}}}_{\mathrm{I}\Gamma} : \widehat{W}_I^{\Gamma} \to \widehat{W}_I^{\mathrm{I}},\ \underline{\underline{\widehat{Q}}}_{\Gamma\mathrm{I}} : \widehat{W}_I^{\mathrm{I}} \to \widehat{W}_I^{\Gamma} \text{ and } \underline{\underline{\widehat{Q}}}_{\Gamma\Gamma} : \widehat{W}_I^{\Gamma} \to \widehat{W}_I^{\Gamma} \quad (9.11)$$

while

$$\underline{\underline{Q}}_{\mathrm{II}} : W_I^{\mathrm{I}} \to W_I^{\mathrm{I}},\ \underline{\underline{Q}}_{\mathrm{I}\Gamma} : W_I^{\Gamma} \to W_I^{\mathrm{I}},\ \underline{\underline{Q}}_{\Gamma\mathrm{I}} : W_I^{\mathrm{I}} \to W_I^{\Gamma} \text{ and } \underline{\underline{Q}}_{\Gamma\Gamma} : W_I^{\Gamma} \to W_I^{\Gamma} \quad (9.12)$$

All this leads to

$$\underline{\underline{Q}} = \begin{pmatrix} \underline{\underline{\widehat{Q}}}_{\mathrm{II}} & \underline{\underline{\widehat{Q}}}_{\mathrm{I}\Gamma} \underline{\underline{\widehat{a}}}^{-1} \\ \underline{\underline{\widehat{a}}} \underline{\underline{\widehat{Q}}}_{\Gamma\mathrm{I}} & \underline{\underline{\widehat{a}}} \underline{\underline{\widehat{Q}}}_{\Gamma\Gamma} \underline{\underline{\widehat{a}}}^{-1} \end{pmatrix} \quad (9.13)$$

Equation Section 10

## 10.- DVS PARALLEL FORMULATION OF THE PROBLEM

In this Section, $\underline{\underline{M}}$ and $\underline{\underline{\widehat{M}}}$ stand for the linear transformations $\underline{\underline{\widehat{M}}} : \widehat{W} \to \widehat{W}$ and $\underline{\underline{M}} : W_I \to W_I$, respectively. Moreover, assume

$$\begin{aligned} &i)\ \underline{\underline{\widehat{M}}} \text{ is non-singular, and} \\ &ii)\ \underline{\underline{M}} \sim \underline{\underline{\widehat{M}}} \end{aligned} \quad (10.1)$$

**DEFINITION 10.1**

Consider the equation

$$\underline{\underline{\widehat{M}}}\,\underline{\widehat{u}} = \underline{\widehat{f}} \tag{10.2}$$

where $\underline{\widehat{u}} \in \widehat{W}$ and $\underline{\widehat{f}} \in \widehat{W}$. In what follows, we refer to Eq.(10.2) as the *'original sequential formulation'*.

**DEFINITION 10.2**

Consider the equation

$$\underline{\underline{M}}\,\underline{u} = \underline{f} \text{ and } \underline{\underline{j}}\,\underline{u} = \underline{0} \tag{10.3}$$

where $\underline{u} \in W$ and $\underline{f} \in W_I$. In what follows, we refer to Eq.(10.3) as the *'DVS-DDM formulation'*.

**LEMMA 10.1**

When $\underline{f} \sim \underline{\widehat{f}}$, then $\underline{\widehat{u}} \in \widehat{W}$ and $\underline{u} \in W_I$ satisfy the *original sequential formulation* and the *DVS-DDM formulation*, respectively, if and only if,

$$\underline{u} \sim \underline{\widehat{u}} \tag{10.4}$$

<u>Proof</u>. We need to prove

$$\left.\begin{array}{l} \underline{\underline{\widehat{M}}}\,\underline{\widehat{u}} = \underline{\widehat{f}} \\ and \\ \underline{\underline{M}}\,\underline{u} = \underline{f} \end{array}\right\} \Leftrightarrow \underline{\widehat{u}} = \underline{\underline{\widehat{a}}}^{-1}\underline{u} \tag{10.5}$$

Introducing the assumptions $\underline{\underline{M}} \sim \underline{\underline{\widehat{M}}}$ and $\underline{f} \sim \underline{\widehat{f}}$

$$\left.\begin{array}{l} \underline{\underline{\widehat{M}}}\,\underline{\widehat{u}} = \underline{\widehat{f}} \\ and \\ \underline{\underline{\widehat{a}}}\,\underline{\underline{\widehat{M}}}\,\underline{\underline{\widehat{a}}}^{-1}\underline{u} = \underline{\underline{\widehat{a}}}\,\underline{\widehat{f}} \end{array}\right\} \Leftrightarrow \underline{\widehat{u}} = \underline{\underline{\widehat{a}}}^{-1}\underline{u} \tag{10.6}$$

We observe that

$$\left.\begin{array}{l} \underline{\underline{\widehat{M}}}\,\underline{\widehat{u}} = \underline{\widehat{f}} \\ and \\ \underline{\underline{\widehat{a}}}\,\underline{\underline{\widehat{M}}}\,\underline{\underline{\widehat{a}}}^{-1}\underline{u} = \underline{\underline{\widehat{a}}}\,\underline{\widehat{f}} \end{array}\right\} \Leftrightarrow \left.\begin{array}{l} \underline{\widehat{u}} = \underline{\underline{\widehat{M}}}^{-1}\underline{\widehat{f}} \\ and \\ \underline{\underline{\widehat{a}}}^{-1}\underline{u} = \underline{\underline{\widehat{M}}}^{-1}\underline{\widehat{f}} \end{array}\right\} \Leftrightarrow \left\{\underline{\widehat{u}} = \underline{\underline{\widehat{a}}}^{-1}\underline{u}\right\} \tag{10.7}$$

**THEOREM 10.1**

When $\underline{f} \sim \underline{\hat{f}}$, then $\underline{\hat{u}} \in \widehat{W}$ and $\underline{u} \in W$ satisfy the *original sequential formulation* and the *DVS-DDM formulation*, respectively, if and only if, $\underline{u} \in W_I$ and

$$\underline{u} \sim \underline{\hat{u}} \tag{10.8}$$

Proof. To prove this Theorem, we observe that the $\underline{\underline{j}}\underline{u} = \underline{0} \Leftrightarrow \underline{u} \in W_I$ and then apply Lemma 10.1.

Equation Section 11

## 11. GENERAL SCHUR-COMPLEMENT DECOMPOSITIONS

Here all matrices considered are defined on the whole *derived-vector-space* $W$ and, when applied to any vector of $W$, they yield *derived-vectors*, i.e., $\underline{\underline{B}}: W \to W$. The symbol $\underline{\underline{B}}$ will be used generically for any such a matrix.

Furthermore, the symbols $N\left(\underline{\underline{B}}\right)$ and $N\left(\underline{\underline{B}}\right)^{\perp}$ stand for the *null-subspace* of $\underline{\underline{B}}$ and its orthogonal complement, respectively. Next, the operators $\underline{\underline{a}}_{\underline{\underline{B}}}: W \to W$ and $\underline{\underline{j}}_{\underline{\underline{B}}}: W \to W$, which are defined by

$$\underline{\underline{a}}_{\underline{\underline{B}}} \equiv Proj_{N\left(\underline{\underline{B}}\right)^{\perp}} \text{ and } \underline{\underline{j}}_{\underline{\underline{B}}} \equiv \underline{\underline{I}} - \underline{\underline{a}}_{\underline{\underline{B}}} \tag{11.1}$$

are introduced.

**Definition 11.1.** The *pseudo-inverse* of $\underline{\underline{B}}$, $\underline{\underline{B}}^{\sim 1}: \left(\underline{\underline{B}}W\right) \to N\left(\underline{\underline{B}}\right)^{\perp} \subset W$, is defined for every $\underline{w} \in \underline{\underline{B}}W$ by the conditions:

$$\underline{v} = \underline{\underline{B}}^{\sim 1}\underline{w} \Leftrightarrow \underline{\underline{B}}\underline{v} = \underline{w} \text{ and } \underline{\underline{j}}_{\underline{\underline{B}}}\underline{v} = 0 \tag{11.2}$$

REMARK 11.1. The *pseudo-inverse* of $\underline{\underline{B}}$, $\underline{\underline{B}}^{\sim 1}: \left(\underline{\underline{B}}W\right) \to N\left(\underline{\underline{B}}\right)^{\perp} \subset W$, is well-defined, i.e., it exists (whenever $\underline{w} \in \underline{\underline{B}}W$) and it is unique.

We now consider two subsets of *derived-nodes*, $M \subset X$ and $N \subset X$, such that

$$W(M) + W(N) \supset N\left(\underline{\underline{B}}\right)^{\perp} \textit{ and } W(M) \perp W(N) \tag{11.3}$$

Under the assumption that Eq. is fulfilled, the following notation is introduced:

$$\sigma_{\mathrm{NN}}^{\mathrm{MM}}\left(\underline{\underline{B}}\right) \equiv \left\{ \underline{\underline{B}}_{\mathrm{NN}} - \underline{\underline{B}}_{\mathrm{NM}} \left(\underline{\underline{B}}_{\mathrm{MM}}\right)^{\sim 1} \underline{\underline{B}}_{\mathrm{MN}} \right\} \tag{11.4}$$

We observe that $\sigma_{\mathrm{NN}}^{\mathrm{MM}}\left(\underline{\underline{B}}\right)$ is a linear operator on $W$; i.e., $\sigma_{\mathrm{NN}}^{\mathrm{MM}}\left(\underline{\underline{B}}\right): W \to W$, although $\sigma_{\mathrm{NN}}^{\mathrm{MM}}\left(\underline{\underline{B}}\right)$ depends on $\underline{\underline{B}}$ in a non-linear fashion.

**Theorem 11.1.** Assume that Eq.(11.3) is satisfied and let $\underline{v} \equiv \underline{v}_{\mathrm{M}} + \underline{v}_{\mathrm{N}} = \underline{\underline{B}}^{\sim 1} \underline{w}$, then

$$\sigma_{\mathrm{NN}}^{\mathrm{MM}}\left(\underline{\underline{B}}\right) \underline{v}_{\mathrm{N}} = \underline{w}_{\mathrm{N}} - \underline{\underline{B}}_{\mathrm{NM}} \left(\underline{\underline{B}}_{\mathrm{MM}}\right)^{\sim 1} \underline{w}_{\mathrm{M}}, \; \underline{\underline{j}}_{B} \underline{v} = 0 \tag{11.5}$$

and

$$\underline{v}_{\mathrm{M}} = \left(\underline{\underline{B}}_{\mathrm{MM}}\right)^{\sim 1} \left(\underline{w}_{\mathrm{M}} - \underline{\underline{B}}_{\mathrm{MN}} \underline{v}_{\mathrm{N}}\right) \tag{11.6}$$

Proof. A more explicit form of equation $\underline{\underline{B}}\underline{v} = \underline{w}$ is

$$\begin{pmatrix} \underline{\underline{B}}_{\mathrm{MM}} \underline{v}_{\mathrm{M}} + \underline{\underline{B}}_{\mathrm{MN}} \underline{v}_{\mathrm{N}} \\ \underline{\underline{B}}_{\mathrm{NM}} \underline{v}_{\mathrm{M}} + \underline{\underline{B}}_{\mathrm{NN}} \underline{v}_{\mathrm{N}} \end{pmatrix} = \begin{pmatrix} \underline{w}_{\mathrm{M}} \\ \underline{w}_{\mathrm{N}} \end{pmatrix} \tag{11.7}$$

Then, the lead to the proof of this Theorem is the standard Schur complement decomposition. The only subtlety is the need of using *pseudo-inverses* where they occur, and this is shown in the Appendix.

If we write

$$\underline{\underline{B}}^{\sim 1} \equiv \begin{Bmatrix} \left(\underline{\underline{B}}^{\sim 1}\right)_{\mathrm{MM}} & \left(\underline{\underline{B}}^{\sim 1}\right)_{\mathrm{MN}} \\ \left(\underline{\underline{B}}^{\sim 1}\right)_{\mathrm{NM}} & \left(\underline{\underline{B}}^{\sim 1}\right)_{\mathrm{NN}} \end{Bmatrix} \tag{11.8}$$

Then, it is seen that

$$\left(\underline{\underline{B}}^{\sim 1}\right)_{\mathrm{MM}} = \left(\underline{\underline{B}}_{\mathrm{MM}}\right)^{\sim 1} + \left(\underline{\underline{B}}_{\mathrm{MM}}\right)^{\sim 1} \underline{\underline{B}}_{\mathrm{MN}} \sigma_{\mathrm{NN}}\left(\underline{\underline{B}}\right)^{\sim 1} \underline{\underline{B}}_{\mathrm{NM}} \left(\underline{\underline{B}}_{\mathrm{MM}}\right)^{\sim 1}, \quad \left(\underline{\underline{B}}^{\sim 1}\right)_{\mathrm{MN}} = -\left(\underline{\underline{B}}_{\mathrm{MM}}\right)^{\sim 1} \underline{\underline{B}}_{\mathrm{MN}} \sigma_{\mathrm{NN}}\left(\underline{\underline{B}}\right)^{\sim 1}$$
$$\left(\underline{\underline{B}}^{\sim 1}\right)_{\mathrm{NM}} = -\sigma_{\mathrm{NN}}\left(\underline{\underline{B}}\right)^{\sim 1} \underline{\underline{B}}_{\mathrm{NM}} \left(\underline{\underline{B}}_{\mathrm{MM}}\right)^{\sim 1}, \qquad \left(\underline{\underline{B}}^{\sim 1}\right)_{\mathrm{NN}} = \sigma_{\mathrm{NN}}\left(\underline{\underline{B}}\right)^{\sim 1}$$

(11.9)Equation Section 12

## 12.- SOLVING WELL-POSED BOUNDARY-VALUE-PROBLEMS

Consider the equation

$$\underline{\underline{M}}\,\underline{u} = \underline{f} \text{ and } \underline{\underline{j}}\,\underline{u} = \underline{0} \tag{12.1}$$

where

$$\underline{\underline{M}} = \begin{pmatrix} \underline{\underline{\widehat{M}}}_{\Pi} & \underline{\underline{\widehat{M}}}_{\Pi\Gamma}\underline{\underline{\widehat{a}}}^{-1} \\ \underline{\underline{\widehat{a}}}\underline{\underline{\widehat{M}}}_{\Gamma\Pi} & \underline{\underline{\widehat{a}}}\underline{\underline{\widehat{M}}}_{\Gamma\Gamma}\underline{\underline{\widehat{a}}}^{-1} \end{pmatrix} \qquad (12.2)$$

Then, the Schur-complement formulation of this problem is

$$\underline{\underline{\widehat{a}}}\left(\underline{\underline{\widehat{M}}}_{\Gamma\Gamma} - \underline{\underline{\widehat{M}}}_{\Gamma\Pi}\left(\underline{\underline{\widehat{M}}}_{\Pi}\right)^{-1}\underline{\underline{\widehat{M}}}_{\Pi\Gamma}\right)\underline{\underline{\widehat{a}}}^{-1}\underline{u}_{\Gamma} = \underline{\underline{\widehat{a}}}\left(\underline{\widehat{f}}_{\Gamma} - \underline{\underline{\widehat{M}}}_{\Gamma\Pi}\left(\underline{\underline{\widehat{M}}}_{\Pi}\right)^{-1}\underline{\widehat{f}}_{\Pi}\right), \; \underline{\underline{j}}\underline{u}_{\Gamma} = 0 \qquad (12.3)$$

and

$$\underline{u}_{\Pi} = \left(\underline{\underline{\widehat{M}}}_{\Pi}\right)^{-1}\left(\underline{\widehat{f}}_{\Pi} - \underline{\underline{\widehat{M}}}_{\Pi\Gamma}\underline{\underline{\widehat{a}}}^{-1}\underline{u}_{\Gamma}\right) \qquad (12.4)$$

Then,

$$\underline{u} = \underline{u}_{\Pi} + \underline{u}_{\Gamma} \qquad (12.5)$$

## 13.- CONCLUSIONS Equation Section 13

Today, there are two versions of the DVS approach to DDM (DVS-DDM), the original DVS introduced in [28, 29] and the Enhanced-DVS (EDVS-DDM), here presented for the first time.

The main advantage of DVS approaches, over canonical approaches, is that they yield block-diagonal matrices whenever they are applied.

The Enhanced-DVS represents a fundamental improvement of the DVS methodology because of its outstanding simplicity and generality, which enlarges DVS' applicability very much. The original DVS requires building a special matrix and such a construction is impossible in many cases. Even when it is feasible such a requirement significantly limits the effectiveness of the DVS-DDM approach.

The Enhanced-DVS, on the other hand, directly uses the original matrix and does not require constructing a special matrix. Therefore, EDVS is applicable to all linear problems. Hence, it represents a fundamental improvement of the DVS methodology. Indeed, because of its outstanding simplicity and generality, "Enhanced-DVS" is a suitable setting for the *domain decomposition methods* of the future.

## ACKNOWLEDGEMENTS

I here acknowledge that this paper is part of a long line of research that we, my research group and myself, have been developing. However, the purely theoretical material contained in it, is based on an idea I had in January of this year (2021) and have developed in solitude.

Special thanks are also expressed to Ernesto Rubio-Acosta who revised the first draft of the article, and to Iván Contreras-Trejo for his technical discussions.

Equation Section 20

# APPENDIX

## A1. SET THEORY

### DEFINITION A.1.

*Let* $\mathrm{E}$ *be a set, while* $\{\mathrm{E}_1,\ldots,\mathrm{E}_M\}$ *is a family of subsets of* $\mathrm{E}$. *Then the family* $\{\mathrm{E}_1,\ldots,\mathrm{E}_M\}$ *is said to decompose* $\mathrm{E}$ *when*

$$\mathrm{E}=\bigcup_{\alpha=1}^{\alpha=M}\mathrm{E}_\alpha \text{ and } \varnothing=\mathrm{E}_\alpha\cap\mathrm{E}_\beta,\ \forall\alpha\neq\beta \tag{20.1}$$

In particular, when $\mathrm{A}\subset\mathrm{E}$ and $\mathrm{B}\subset\mathrm{E}$ are such that

$$\mathrm{E}=\mathrm{A}\cup\mathrm{B} \text{ and } \varnothing=\mathrm{A}\cap\mathrm{B} \tag{20.2}$$

then, the pair $\{\mathrm{A},\mathrm{B}\}$ is said to *decompose* $\mathrm{E}$.